\documentclass[apj]{emulateapj}
\usepackage{graphicx,natbib,amsfonts,amssymb,rotating,lscape}
\usepackage{apjfonts}
\usepackage{color}
\usepackage{lscape}     %if use preprint, please de-comment \volnopage{} too.
\usepackage{graphicx}
\usepackage{natbib}
\newcommand{\feii}{Fe {\sc ii}\ }
\newcommand{\ha}{\ifmmode H\alpha \else H$\alpha$\ \fi}
\newcommand{\hb}{\ifmmode H\beta \else H$\beta$\ \fi}

\newcommand{\oiii}{[O {\sc iii}]\ }
\newcommand{\hbb}{\ifmmode H\beta^{b} \else H$\beta^{b}$\ \fi}
\newcommand{\hbn}{\ifmmode H\beta^{n} \else H$\beta^{n}$\ \fi}
\newcommand{\ergs}{\ifmmode {\rm erg\ s}^{-1} \else erg s$^{-1}$\ \fi}
\newcommand{\civ}{C {\sc iv}\ }
\newcommand{\ciii}{C {\sc iii]}\ }
\newcommand{\mgii}{Mg {\sc ii}\ }
\newcommand{\nv}{N {\sc v}\ }
\newcommand{\kms}{\ifmmode {\rm km\ s}^{-1} \else km s$^{-1}$\ \fi}
\newcommand{\msun}{\ifmmode M_{\odot} \else $M_{\odot}$\ \fi}
\newcommand{\lv}{\ifmmode L_{\lambda}(5100\AA) \else $L_{\lambda}(5100\AA)$\ \fi}

\begin{document}
\title{Spectral variability of FIRST bright QSOs with SDSS observations}
\author{Wei-Hao Bian$^{1,2}$, Li Zhang$^{1}$, Richard Green$^{2}$, Chen Hu$^{3}$\\
$^{1}$Department of Physics and Institute of Theoretical Physics,
Nanjing Normal University, Nanjing
210097, China\\
$^{2}$ LBT Observatory, AZ, 85721, USA \\
$^{3}$ Key Laboratory for Particle Astrophysics, Institute of High Energy Physics,
Chinese Academy of Sciences, Beijing 100039, China \\
} \shorttitle{QSO spectral variability} \shortauthors{Bian, et al.}
\begin{abstract}
For some samples, it has been shown that spectra of QSOs with low
redshift are bluer during their brighter phases. For the FIRST
bright QSO sample, we assemble their spectra from SDSS DR7 to
investigate variability between the spectra from White et al. (2000)
and from the SDSS for a long rest-frame time-lag, up to 10 years.
There are 312 radio loud QSOs and 232 radio quiet QSOs in this
sample, up to $z \sim 3.5$. With two-epoch variation, it is found
that spectra of half of the QSOs appear redder during their brighter
phases. There is no obvious difference in slope variability between
sub-samples of radio quiet and radio loud QSOs. This result implies
that the presence of a radio jet does not affect the slope
variability on 10-year timescales. The arithmetic composite
difference spectrum for variable QSOs is steep at blueward of $\sim$
2500\AA. The variability for the region blueward of 2500 \AA\ is
different to that for the region redward of 2500 \AA.

\end{abstract}
\keywords{galaxies: active --quasars: general -- techniques:
spectroscopic}
\section{Introduction}
Variability is a common phenomenon in Active Galactic Nuclei (AGNs)
and QSOs, from X-ray to radio wavelengths, on timescales from hours
to decades (e.g., Huang et al. 1990; Mushotzky et al. 1993; Bian \&
Zhao 2003a; Breedt et al. 2010). Detection of variability is also a
method to isolate QSO samples from photometric data (e.g., Rengstorf
2004; Wu et al. 2011). The origin of the small-amplitude and
long-time-scale optical variability in AGN is still a question to
debate, and there are several main models: accretion disc
instabilities (e.g. Rees 1984; Kawaguchi et al. 1998), gravitational
lensing (e.g. Hawkins 1996), and star/supernova activity (e.g. Cid
Fernandes, Aretxaga \& Terlevich 1996, Torricelli-Ciamponi et al.
2000), reprocessing of X-rays and reflection of optical light by the
dust(e.g., Breedt et al. 2010).  It is suggested that the predicted
slopes of the structure function for QSOs variability are 0.83,
0.44, and 0.25 for supernovae in the starburst model, instabilities
in the accretion disk, and microlensing, respectively (Kawaguchi et
al. 1998; Hawkins 2002).

There are two main methods to investigate QSOs variability: one is
from photometry, the other is from spectra. With the photometric
method, many correlations are found between photometric variability
and luminosity, rest-frame wavelength, redshift, time-lag,
supermassive lack hole mass (SMBH mass), and Eddington ratio (e.g.,
Huang, et al. 1990; Hawkins 2002; Vanden Berk et al. 2004; de Vries
et al. 2005; Bauer et al. 2009; Ai et al. 2010; Meusinger et al.
2011). Using the photometric data in various optical bands, the
spectral variability can also be studied, suggesting that the QSOs
would be bluer during their brighter phase (e.g., Giveon et al.
1999; Vanden Berk et al. 2004; de Vries et al. 2005; Meusinger et
al. 2011; Gu \& Ai 2011). With large QSO samples, increasing
variability at shorter wavelengths supports accretion disk
instability as the explanation for the structure function (e.g.,
Vanden Berk et al. 2004; de Vries et al. 2005; Bauer et al. 2009).

The advantage of the photometric method is that the photometric
variability amplitude is accurate and that it takes less observing
time compared with the spectroscopic method. However, it just
monitors the flux variability in a few wavelength bands which
usually have contributions from many components, including continuum
and strong emission lines. In order to investigate the spectral
variability in QSOs in more detail, the spectroscopy-based method is
necessary. Wilhite et al. (2005) gave a sample of 315 significantly
variable QSOs from multi-epoch spectroscopic observations of the
SDSS. They found that the average difference spectrum (bright phase
minus faint phase) is bluer than the average single-epoch QSO
spectrum, also suggesting that QSOs are bluer when brighter (see
also Meusinger et al. 2011). Pu et al. (2009) and Bian et al. (2010)
also investigated spectral variability with data from the
Palomer-Green (PG) QSOs spectrophotometrical monitoring projects for
individual QSOs, but with many epochs ($\sim 20-70$). They also
found this result for all these individual QSOs. However, these
objects are very nearby, with low redshifts. Their spectral slopes
are influenced by their host galaxy contributions (e.g., Shen et al.
2011).

There is usually a change in the distribution of radio to optical
flux ratio for QSO sample, at which the population can be divided
into radio loud (RL) QSOs and radio quiet (RQ) QSOs, and its origin
is still a question open to debate (e.g., White et al 2000; Jiang et
al. 2008). It is suggested that the difference between RL and RQ
QSOs is due to their SMBH masses, their SMBH spins, or Eddington
ratios (e.g., Woo \& Urry 2002; Ho 2002; Laor 2003; Bian \& Zhao
2003b). It is interesting to investigate the spectral variation for
RL QSOs and RQ QSOs. For the objects from the FIRST bright QSOs
survey (White et al. 2000), we assembled their spectra from the
Sloan Digital Sky Survey Data Release 7 (SDSS DR7; York et al. 2000;
Abazajian et al. 2009) to investigate the variability between the
spectra from White et al. (2000) and from the SDSS for a long
rest-frame time-lag, up to 10 years. The slope variation is measured
for the two epochs, as well as composite spectra to construct the
ratio and difference spectra.

In Section 2, the sample and data are described.The spectroscopic
data analysis is given in Section 3. Results and discussions are
presented in section 4. Section 5 is our conclusion.

\section{Sample and Data}

\subsection{The FBQS with SDSS spectra}
With the FIRST survey and the Automated Plate Measuring Facility
(APM) catalog of the Palomar Observatory Sky Survey I (POSS-I)
plates, White et al. (2000) presented their FIRST Bright Quasar
Survey (FBQS) catalog with 636 QSOs distributed over 2682 $deg^2$.
There are four criteria to select FBQS candidates: radio-optical
positional coincidence (1".2); E$\le$ 17.8; optical morphology
(stellar-like); POSS color bluer than O-E=2. They do not find a
bimodal distribution in radio to optical flux ratio, and the numbers
of RQ/RL QSOs are comparable. Spectra for the FBQS were obtained at
different observatories: the 3 m Shane telescope at Lick
Observatory, the 2.1 m telescope at Kitt Peak National Observatory,
the 3.5 m telescope at Apache Point Observatory,the $6 \times 1.8 m$
Multiple Mirror Telescope (MMT), and the 10 m Keck II telescope.
These FIRST spectra can be downloaded from the CDS
\footnote{http://cdsarc.u-strasbg.fr/viz-bin/Cat?J/ApJS/126/133}.

The SDSS DR7 (York et al. 2000) contains imaging of almost 11663
$deg^2$ and follow-up spectra for roughly $93\times 10^4$ galaxies
and $12\times 10^4$ quasars. All observations were made at the
Apache Point Observatory in New Mexico, using a dedicated 2.5m
telescope. The SDSS spectra were obtained through 3" fibers. With
these 636 FBQS, we cross-correlate them with the total SpecPhotoALL
table \footnote {http://cas.sdss.org/CasJobs/MyDB.aspx} in the SDSS
DR7. Applying the FBQS criterion of stellar-like morphology, we find
544 objects in SDSS DR7 spectral database with separations less than
1", which is about 86\% of 636 FBQSs. There are 312 RL QSOs and 232
RQ QSOs. Except for several objects, most of the FBQS objects have
only one spectrum in the SDSS DR7 spectral database. For the SDSS
spectra, the observational wavelength coverage is from ~3800 to
9200\AA. For spectra from White et al. (2000), the observational
wavelength coverage can be found in their Table 1, which typically
stops at a little shorter wavelength than the SDSS spectrum in the
red part and starts a little longer than the SDSS spectrum in the
blue part.

\subsection{The subsample of spectrally variable QSOs}

\begin{figure*}
\centering
\includegraphics[height=7cm,angle=-90]{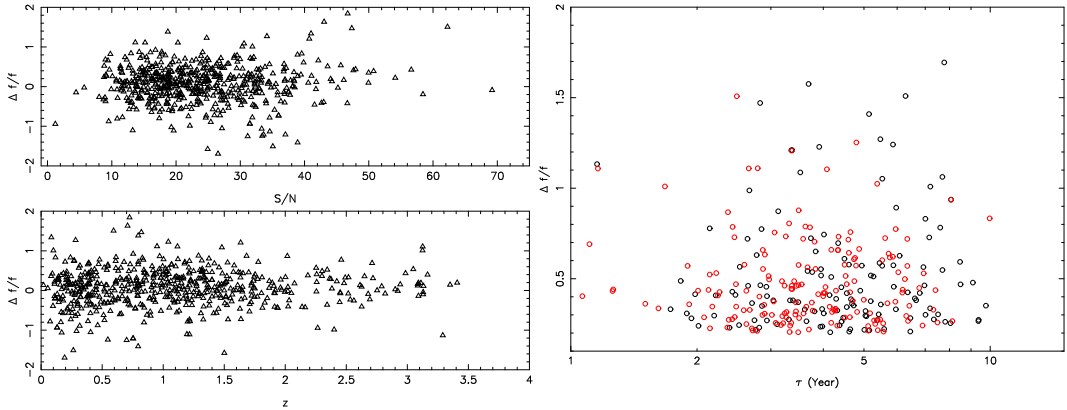}
\includegraphics[height=7cm,angle=-90]{f1b.eps}
\caption{Left panel: the integrated relative flux change versus S/N
from SDSS (top); the integrated relative flux change versus redshift
(bottom). Right panel: the absolute value of integrated relative
flux change versus the rest-frame time-lag for the variable QSOs.
Red circles are  RL QSOs, and black circles are RQ QSOs. }
\end{figure*}

In order to isolate the spectrally variable QSOs from the total
QSOs, we calculate the integrated relative flux change ($\Delta
f/f$). For the two-epoch spectra of a QSO, we calculate the mean
flux ($f$) and the difference flux ($\Delta f$) at every wavelength
(SDSS flux minus FIRST flux). This integrated relative flux change
measures the total relative flux change between the two epochs over
a large optical wavelength range, rather than the variation at just
one wavelength or in one filter. With the stars to calibrate the QSO
flux variability, Wilhite et al. (2005) found that the criterion for
selecting spectrally variable QSOs is that the absolute value of the
integrated $\Delta f/f$ is larger than 0.2-0.1 when the high-S/N
epoch S/N is below 50. Here, we examine two-epoch spectra and find
this criterion of 0.2 is appropriate to select the spectrally
variable QSOs. In the left panel of Figure 1, we give the integrated
relative flux change, $\Delta f/f$, versus the integrated S/N from
SDSS and the redshift. The total numbers of two-epoch measurements
are 242 and 319 for RQ QSOs and RL QSOs. Considering the S/N in the
ratio spectrum, we select QSOs with ratio spectral S/N greater than
5.0. There are 221 RQ QSOs and 269 RL QSOs ratio spectra meeting the
criterion. For the subsample of variable QSOs, the two-epoch numbers
are 142 (58\%) for variable RQ QSOs, and 204 (64\%) for variable RL
QSOs. We define the convention that the SDSS spectrum is in the
brighter phase and the relevant FIRST spectrum is in the fainter
phase if $\Delta f/f$ is positive.

For our variable QSO sample, the rest-frame time-lag is from $\sim
1$ to $\sim 10$ years (right panel in Fig. 1), while the rest-frame
time-lag is less than 1 year for Whihite et al. (their Fig. 7,
2005), less than 2 years for Vanden Berk et al. (2004). We calculate
structure functions at three continuum wavelengths of 5100\AA,
3000\AA, and 1350\AA. However, the data are so dispersed that we
can't give the slopes measurement of the structure functions. The
right panel of Figure 1 is the integrated relative flux change
versus the rest-frame time-lag. For the distribution of $\Delta
f/f$, K-S statistic $d$ is 0.081 with a signification level of 81\%.
Assuming that the RL and RQ QSOs were drawn from the same
population, the probability that we would have observed a K-S
statistics as at least as large as 0.081 is 81\%. There is no
obvious difference of $\Delta f/f$ for variable RL QSOs and RQ QSOs.

\section{spectroscopic analysis}

Our goal is to investigate the spectral variability from the
two-epoch variation for the total QSO sample, as well as the
sub-samples of RL and RQ QSOs. All the observed spectra are
corrected for Galactic extinction using their $A_V$ values, assuming
an extinction curve of Cardelli et al. (1989; IR band; UV band) and
O'Donnell (1994; optical band) with $R_V = 3.1$. The correction for
Galactic extinction has no effect on the slope variation
calculation, however, it would have a large effect on the continuum
flux, as well as the slope of the composite spectrum.

It is common to use the power law formula, $f_\lambda \propto
\lambda ^{\alpha} $ ($f_\upsilon \propto \upsilon ^{-(2+\alpha)} $)
to approximately fit the QSO continuum spectrum. We use "continuum
windows" (in the rest frame), known to be relatively free from
strong emission lines, of 1350-1370, 1450-1470, 1680-1720,
2200-2250, 3790-3810, 3920-3940, 4000-4050, 4200-4230, 5080-5100,
5600-5630, 5970-6000, 6005-6035, 6890-7010\AA\ (Forster et al. 2001;
Vanden Berk et al. 2001). Because of the pseudo-continuum from
UV/optical Fe II (from 2200 to 3800 \AA, from 4400 to 5500\AA), the
Balmer continuum (blueward of $\sim$ 4000\AA), as well as the host
contribution for low luminosity QSOs, there is some difficulty in
finding the ideal "continuum windows". At the same time, for FBQS
spectra from White et al. (2000), there are two obvious strong
atmospheric absorptions around $\sim$ 6900 \AA\ and $\sim$ 7600\AA\
in the observational frame. These two regions are excluded in the
fitting. Considering the errors in both coordinates, we use the
linear regression algorithm, $\it{fitexy}$, to parameterize the
power-law fit (Press et al. 1992, p. 660).

For most objects, we have one spectrum from SDSS, $f_\lambda^{SDSS}
\propto \lambda ^{\alpha_{SDSS}} $, one spectrum from FBQS of White
et al. (2000), $f_\lambda^{FIRST} \propto \lambda ^{\alpha_{FIRST}}
$, as well as the ratio spectrum from SDSS/FBQS flux ratio,
$f_\lambda^{SDSS} /f_\lambda^{FIRST}\propto \lambda ^{\Delta \alpha}
$. With three power-law fits for three spectra, we can obtain three
slopes, three intercepts and their errors. In the spectrum of the
SDSS/FBQS flux ratio, the emission line contribution would vanish or
be decreased (Wilhite et al. 2005). Therefore, the slope variation
between the SDSS spectrum and the FBQS spectrum is adopted to be the
power-law fit to their spectral flux ratio, rather than the slope
difference from the direct SDSS and FBQS spectral fits (Figure 2).

\begin{figure*}
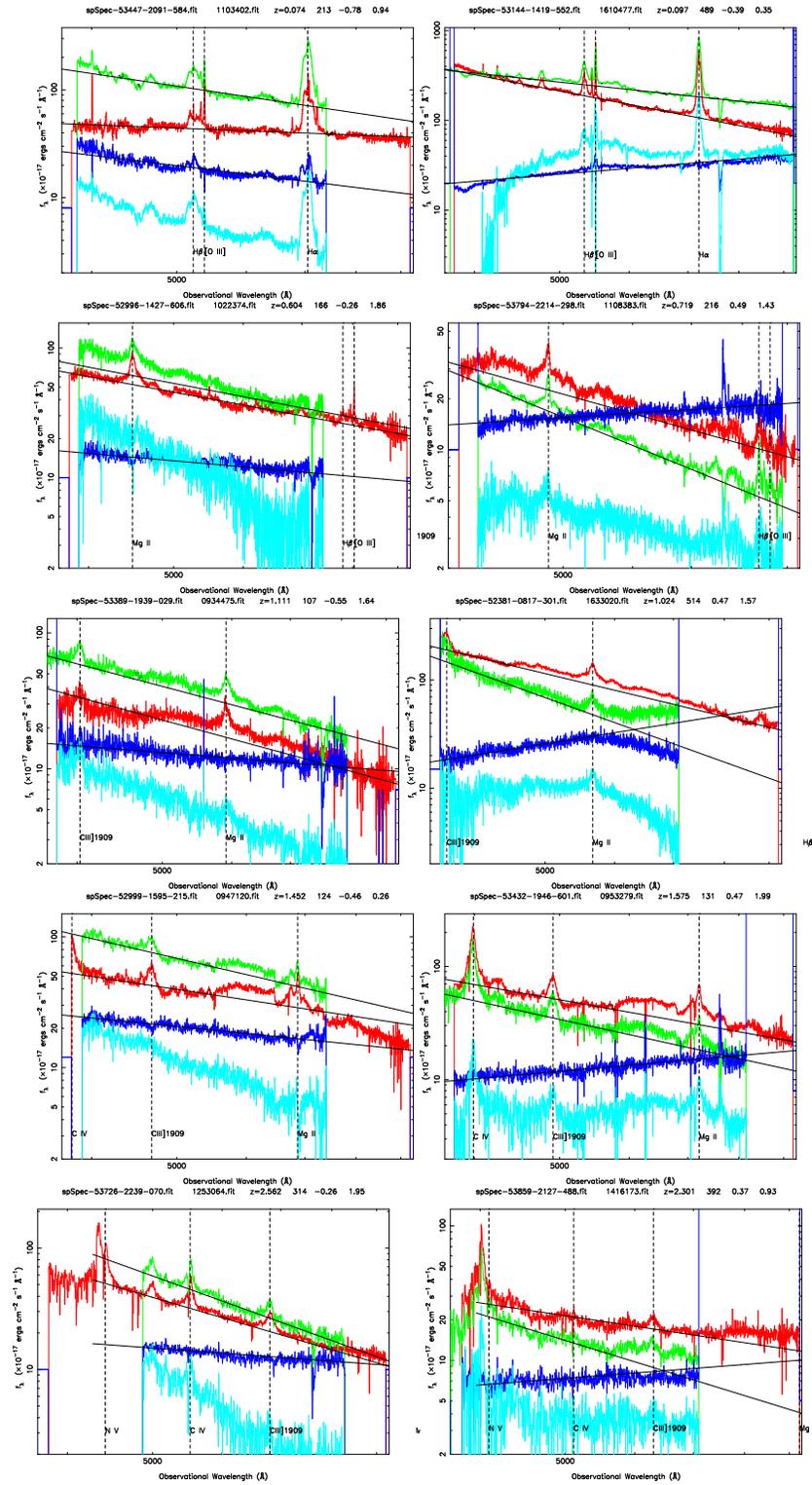

\centering
\includegraphics[width=4cm,angle=-90]{f2a.eps}
\includegraphics[width=4cm,angle=-90]{f2b.eps}\\
\includegraphics[width=4cm,angle=-90]{f2c.eps}
\includegraphics[width=4cm,angle=-90]{f2d.eps}\\
\includegraphics[width=4cm,angle=-90]{f2e.eps}
\includegraphics[width=4cm,angle=-90]{f2f.eps}\\
\includegraphics[width=4cm,angle=-90]{f2g.eps}
\includegraphics[width=4cm,angle=-90]{f2h.eps}\\
\includegraphics[width=4cm,angle=-90]{f2i.eps}
\includegraphics[width=4cm,angle=-90]{f2j.eps}
\vspace{-0.1cm} \caption{Examples of spectral variability. The red
curve is the spectrum from SDSS; the green curve is from the FBQS
spectrum by White et al. (2000); the blue curve is the ratio of the
bright-phase spectrum to the faint-phase spectrum (flux scaled for
clarity); the cyan cure is their different spectrum (flux scaled for
clarity). The solid lines are the best linear fits at "continuum
windows" (black points). The left figures show that QSOs become
bluer during their brighter phase, while the right figures show a
another trend, i.e., redder during their brighter phase. At the top
of each panel, we list the fits file names of the SDSS and FIRST
spectra, redshift, the number, the integrated relative flux change,
and logarithm of the radio loudness. In each panel, we also show the
main emission lines (dot vertical straight lines), such as \ha, \hb,
\oiii, \mgii, \ciii, \civ, \nv.}
\end{figure*}

Figure 2 is an example showing QSOs two-epoch variability for 5
pairs of QSOs with different redshifts (from top to bottom, about
0.1, 0.6, 1.0, 1.5, 2.5). The red curve is for the SDSS spectrum,
the green is for the FBQS spectrum from White et al. (2000); the
blue curve is for the spectral flux ratio of the bright-phase
spectrum to the faint-phase spectrum; the cyan cure is their
different spectrum. The solid lines are the  best linear fits at
"continuum windows" (black points) by $\it{fitexy}$ . The left
panels show the QSOs that become bluer during the brighter phase,
while the right panels show another trend, i.e., redder during the
brighter phase.

\section{Results and Discussions}

\subsection{The distributions of spectral slope and slope variation}

\begin{figure*}
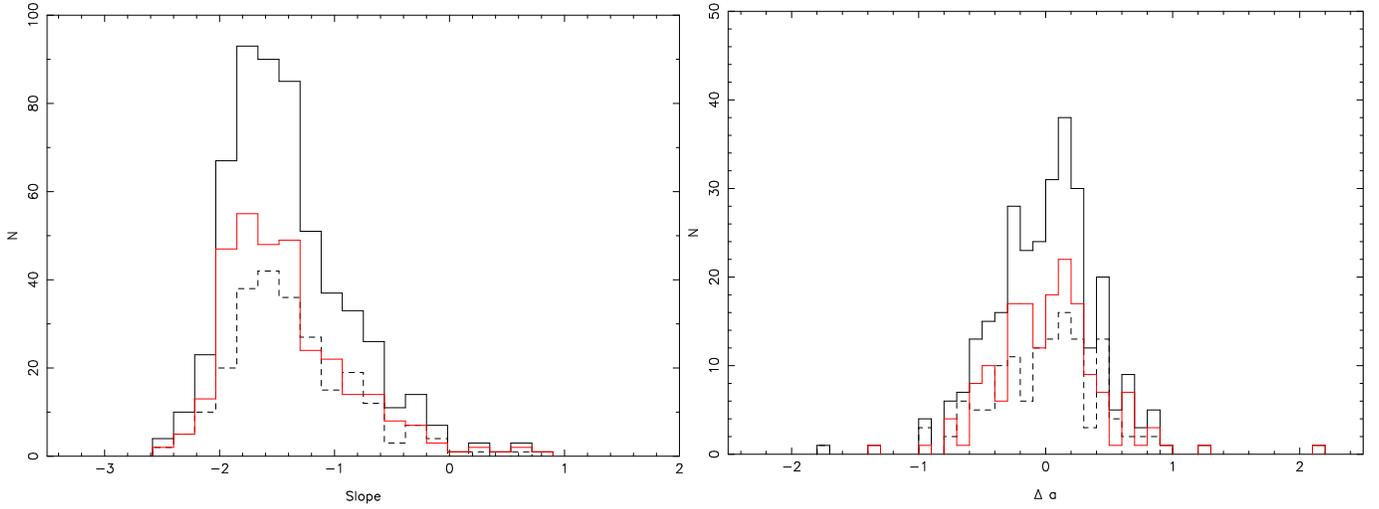

\includegraphics[height=9cm,angle=-90]{f3a.eps}
\includegraphics[height=9cm,angle=-90]{f3b.eps}
\caption{Left: the spectral slope distribution for all SDSS QSOs in
the sample. Right: the spectral slope variation for variable QSOs
(the slope of brighter phase minus the slope of fainter phase). The
black solid curve is for the total sample; the black dash line is
for RQ QSOs; the red solid curve is for RL QSOs.}
\end{figure*}

Figure 3 is the spectral slope distributions for SDSS QSOs (left
panel) and the slope variation distribution (right panel). The mean
of the slope distribution of SDSS QSOs is $-1.41$ with a standard
deviation of 0.59. It is $-1.39\pm 0.56$, $-1.41\pm 0.51$,
respectively for RQ QSOs and RL QSOs from SDSS. In checking the
spectra with the reddest slopes, it is found that these are mainly
QSOs with obvious host galaxy contribution. For the slope variation
distribution in the subsample of variable QSOs, the mean slope
variation (the slope of brighter phase minus the slope of the
fainter phase) is $-0.003$ with a standard deviation of 0.42. It is
$-0.03\pm 0.42$, $0.02\pm 0.42$, respectively for variable RQ QSOs
and RL QSOs.

In Figure 4, we give the slope variation versus the integrated
relative flux change $\Delta f/f$ for selected variable QSOs. About
half QSOs get redder during their brighter phases. Selecting for
$\Delta f/f$ larger than 1.0, this is still the case. For the
distribution of $\Delta \alpha$, K-S statistic $d$ is 0.066 with a
signification level of 90\%. Assuming that the RL and RQ QSOs were
drawn from the same population, the probability that we would have
observed a K-S statistics as at least as large as 0.066 is 90\%.

\subsection{Relation between the slope variation and the continuum flux ratio}

For the UV/optical spectrum, clean determinations of continuum
fluxes are mainly at three wavelengths from 5100\AA, 3000\AA, and
1350\AA, which are used in the mass calculation (et al., Shen et al.
2011). These three wavelengths of 5100\AA, 3000\AA, and 1350\AA\ are
selected to discuss the relationship between the slope variation and
the continuum flux ratio. The continuum flux ratio is the flux at
the brighter epoch divided by the flux at the fainter epoch at the
specified rest wavelength. In order to exclude the limits of
observed spectral range, we consider the slope variation at a
wavelength of 5100\AA\ for variable QSOs with $z<0.7$, at a
wavelength of 3000\AA\ for variable QSOs with $0.7<z<1.9$, and at a
wavelength of 1350\AA\ for variable QSOs with $z>1.9$.

In Figure 5, we give the relation between the slope variation and
the continuum flux ratio at 5100\AA\ for low redshift selected
variable QSOs ($z <0.7$) (top left, top right panel); at 3000 \AA\
($0.7 < z < 1.9$, bottom left panel); at 1350 \AA\ ($z > 1.9$,
bottom right panel). For the top right panel, it is done for the
\oiii flux correction in FIRST spectra. We did not find strong
correlations between the slope variation and the continuum flux
ratio, the Spearman coefficients are -0.03, -0.07, -0.10, -0.25, and
the probabilities of null hypothesis are 0.75, 0.55, 0.24, 0.14,
respectively (Figure 5, from left to right and from top to bottom).

It is also clear that the slope variation of half the points is
positive when the flux ratio is larger than zero in all different
subsamples, i.e., about half of the objects become redder during
their brighter phases. The red points are for RL QSOs and the black
points are for RQ QSOs. In low redshift QSOs, the host would
contribute in the observed object spectrum and make it complex
(e.g., Shen et al. 2011). For the low luminosity QSO sub-sample (\lv
< $10^{45}$ \ergs), it is still the case. The top right panel in
Figure 5 is the same relation but with the recalibration for FIRST
spectra from White et al. (2000) from the \oiii 5007 flux (see
section 4.4).

\begin{figure*}
\centering
\includegraphics[height=9cm,angle=-90]{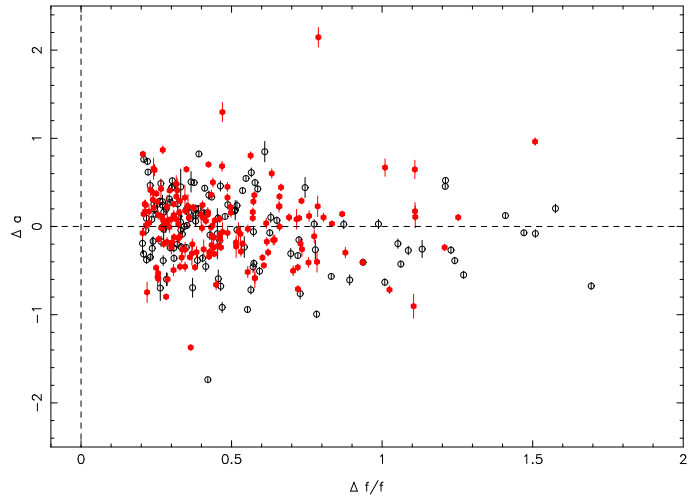}
\caption{The slope variability versus the integrated relative flux
change for selected variable QSOs. Black circles are radio quiet
QSOs and red circles are radio loud QSOs. }
\end{figure*}

\begin{figure*}
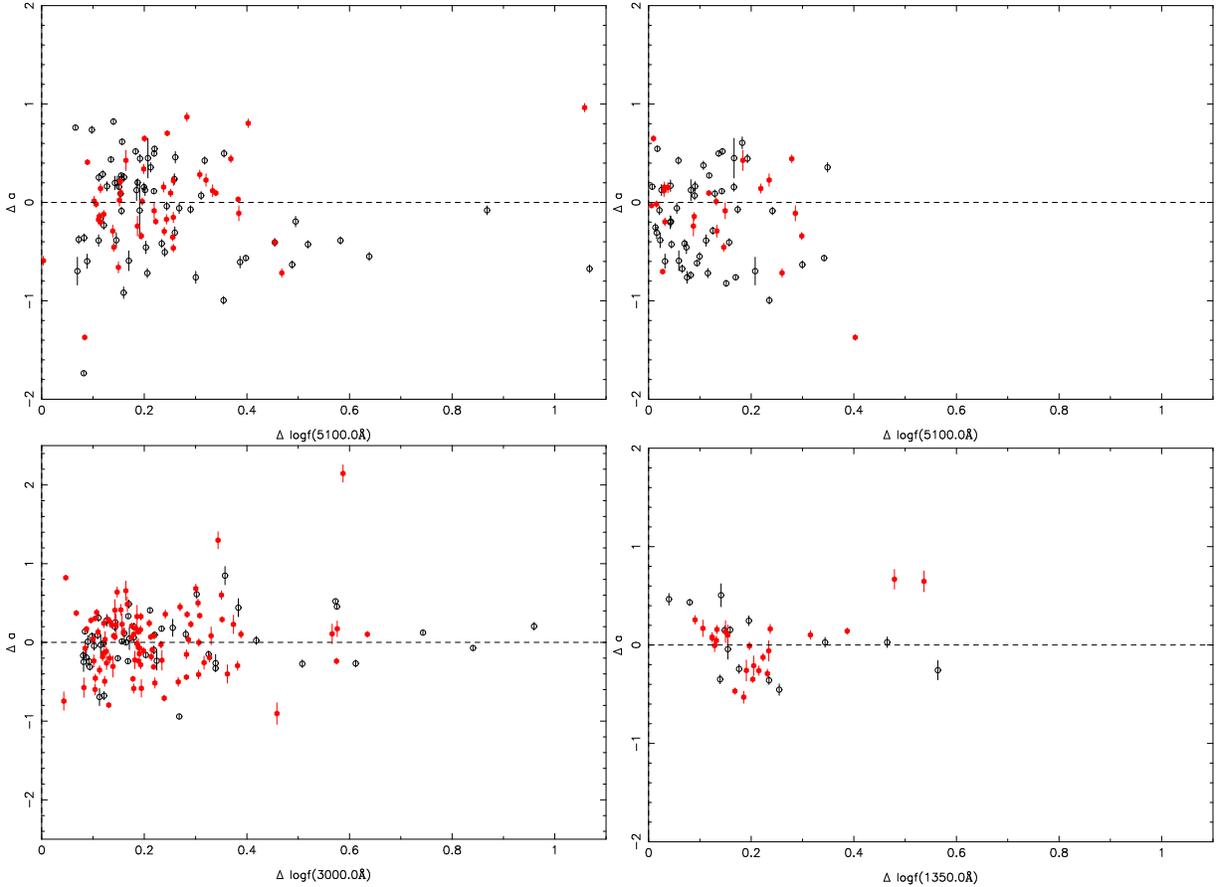

\centering
\includegraphics[height=8cm,angle=-90]{f5a.eps}
\includegraphics[height=8cm,angle=-90]{f5b.eps}
\includegraphics[height=8cm,angle=-90]{f5c.eps}
\includegraphics[height=8cm,angle=-90]{f5d.eps}
\caption{The slope variability versus the continuum flux ratio at
5100\AA\ for low redshift selected variable QSOs ($z <0.7$) (top
left panel, top right panel); at 3000 \AA\  ($0.7 < z < 1.9$, bottom
left panel); at 1350 \AA\ ($z > 1.9$, bottom right panel). Black
circles are radio quiet QSOs and red circles are radio loud QSOs.
For the top right panel, it is done for the \oiii flux correction in
FIRST spectra. The typical error of the continuum flux ratio is 0.2
dex.}
\end{figure*}

\subsection{Composite ratio spectrum}

\begin{figure*}
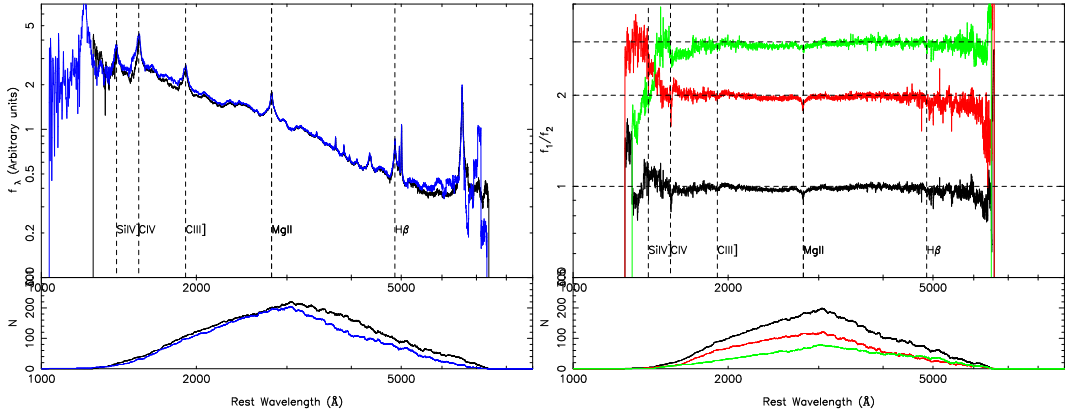

\centering
\includegraphics[height=7cm,angle=-90]{f6a.eps}
\includegraphics[height=7cm,angle=-90]{f6b.eps}
\caption{Left: Geometric composite spectra from SDSS (black curve)
and from FIRST (blue curve), and they are almost the same.  Right:
Composite ratio spectrum for variable QSOs (the flux ratio of
brighter phase spectrum to fainter phase spectrum). All spectra
scaled to flux at 3060\AA. Black curve is for total variable QSOs,
red curve is for variable radio loud QSOs, and green curve is for
variable radio quiet QSOs.}
\end{figure*}

In order to clarify the average slope variability of RL and RQ QSOs,
we calculate their composite ratio spectra. There are two methods to
derive the composite spectrum, one is the median spectrum, which can
preserve the relative fluxes of the emission lines; the other is the
geometric mean spectrum, which can preserve the global continuum
shape. In order to investigate the slope variation, we use the
geometric mean spectrum, $<f_\lambda
>_{gm}=(\prod^{n}_{i=1}f_{\lambda, i})^{1/n}$, where $f_{\lambda,
i}$ is the flux of spectrum number i at wavelength $\lambda$, and n
is the total number of spectra contributing to the bin (e.g., Vanden
Berk et al. 2001). In the left panel of Figure 6, the geometric
composite spectra of our SDSS QSOs and FBQS QSOs are very
consistent. Vanden Berk et al. (2001) suggested the best fitting
windows in the composite spectrum are 1350-1365\AA\ and
4200-4230\AA. Using a power law function to fit the composite
spectrum in the windows of 1450-1470\AA\ and 4200-4320\AA, we find
that their slopes are $-1.41\pm 0.03$ for SDSS spectra and $-1.47\pm
0.03$ for FIRST spectra, which is a little larger than -1.56 by
Vanden Berk et al. (2001) for a large sample of 2204 QSOs from SDSS.

Here we also use that approach to calculate the geometric composite
spectrum for the flux ratio spectrum. The flux ratio is the ratio of
the brighter phase spectrum to the fainter phase spectrum. All
spectra of variable QSOs are scaled to the continuum flux at
3060\AA. In the right panel of Figure 6, we show the geometric mean
composite spectra for the flux ratio spectra for variable RL and RQ
QSOs. The black curve is for all QSOs, the red curve is for RL QSOs,
and the green one is for RQ QSOs. For clarity, the magnitudes are
multiplied by 2 and 3 for RL and RQ QSOs. The geometric mean
flux-ratio spectrum for RL QSOs is almost the same as that for RQ
QSOs. All the geometric mean composite spectra are very flat. This
is consistent with our results in Figures 3, 4, and 5, where about
half of the objects have spectra becoming redder during their
brighter phases.

For the slope variability for individual QSOs, we found that all the
nearby PG QSOs in the sample of Kaspi et al. (2000; for the
reverberation mapping technique) appear bluer during their brighter
phases over multiple-epochs (Pu et al. 2006). However, when just
considering two-epoch, it is not the case (see Fig. 1 in Pu et al.
2006). Using just two-epoch to investigate the slope variability for
an entire ensemble of quasars, Wilhite et al. (2005) suggested that
the QSOs continua are bluer when brighter. Here we directly
calculate the slope variation for every object and find that almost
half the objects appear redder continua when brighter. This paper
uses the analysis framework developed in Pu et al.(2006) on a much
bigger sample of QSOs. The advantages of the current sample are that
it has a roughly equal fraction of RL and RQ objects, and that it
contains a substantial sub-sample of objects that vary strongly in
the optical/UV.

\subsection{Difference of flux calibration between SDSS spectra
and spectra from White et al.}

\begin{figure*}
\centering
\includegraphics[height=9cm,angle=-90]{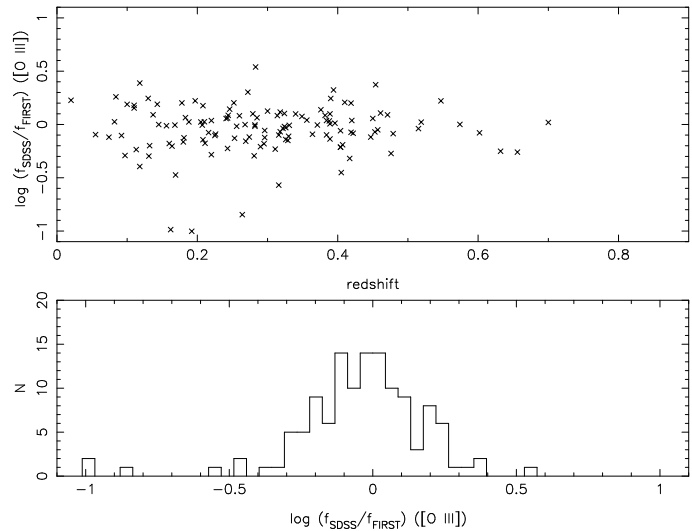}
\caption{Top: the \oiii 5007 \AA\ flux ratio of the SDSS spectra to
White's spectra versus the redshift; Bottom: the distribution of
\oiii flux ratio.}
\end{figure*}

The uncertainty of the slope variation is mainly due to the
selection of "continuum windows" in the power-law fitting. Its mean
error is less than 0.1. For the continuum flux ratio, the
uncertainty is mainly due to the different calibration of SDSS
spectra and spectra from White et al. (2000). For SDSS spectra, with
the calibration from non-variable stars, it is suggested that the
flux correction between High-S/N to low-S/N is about 1.03-1.01
between 4000\AA\ and 9000\AA\ (Wilhite et al. 2005). In logarithm
space, it is 0.004-0.01, or 0.01-0.025 mag. For the spectra from
White et al. (2000), we use the flux of  \oiii 5007\AA\ to estimate
their flux uncertainties.

Considering that \oiii 5007\AA\ emission is coming from narrow line
region (NLRs), spatially extended low-density regions, we assume
that the \oiii fluxes do not vary significantly over several year
timescales and then normalize each FIRST spectrum to have the same
\oiii flux as the SDSS spectrum. Due to the blending of \hb with
\oiii, we use the following steps to calculate the \oiii flux (see
Hu et al. 2008 for details). The first step is subtraction of the
power-law, Balmer continuum, and pseudo-continuum \feii emission.
The second is the multi-Gaussian fit to the \oiii 4959, 5007 \AA\
doublet lines and Hermite-Gaussian for \hb. We excluded objects when
the absorption in ~6900\AA\ and ~7600\AA\ is in the fitting region
of \oiii 5007\AA\ and \hb. The measured flux ratio for \oiii
5007\AA\ is shown in Figure 7. The mean value of the \oiii flux
ratio in logarithm space is $-0.04$ with a standard deviation of
0.23. It suggests that the SDSS flux calibration is consistent with
that for FBQS but with a moderate dispersion.

In the top right panels in Figure 5, we show the slope variation
versus the bright-to-faint flux ratio after considering the \oiii
5007\AA\ correction for the spectra from White et al. (2000). It
maintains the result that almost half of the QSOs are not bluer
during their bright phases. Of course, that correction can't be used
for higher redshift QSOs without observations of \oiii 5007\AA. In
Figure 5, considering the continuum flux ratio larger than 0.23 from
the \oiii recalibration, a little more QSOs with z < 0.7 tend to be
bluer during brighter phases, and there is still half QSOs showing
redder during brighter phases for QSOs with $z > 0.7$.

\subsection{Composite spectrum and the composite difference spectrum}

\begin{figure*}
\centering
\includegraphics[height=12cm,angle=-90]{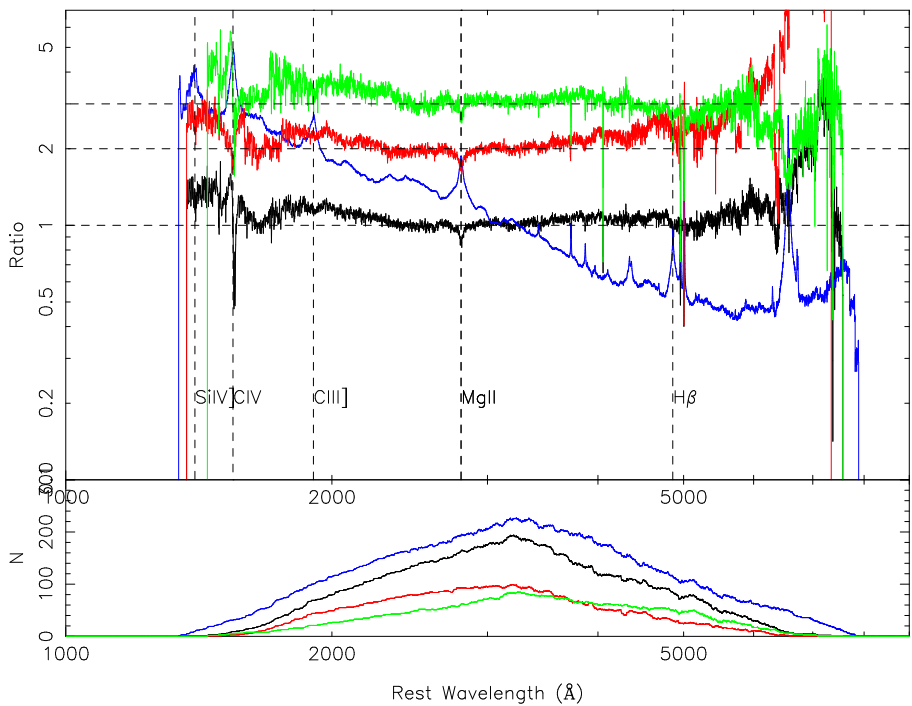}
\caption{Ratio spectrum of the arithmetic composite difference
spectrum to the arithmetic composite spectrum from SDSS (blue
curve). Black curve is for the total variable QSOs, red curve is for
variable RL QSOs (multiplied by 2), green curve is for variable RQ
QSOs (multiplied by 3). The bottom panels show the number
distributions for corresponding different curves.}
\end{figure*}

It is suggested that arithmetic and geometric mean composites have
different properties (Vanden Berk et al. 2001). Because QSO spectra
can be described by power laws, by the definition of geometric mean,
the geometric mean can preserve the average power-law slope. The
arithmetic mean can retain the relative strength of the
non-power-law features, such as emission lines.

For the subsample of variable QSOs, the composite difference
spectrum is calculated from the geometric/arithmetic mean for QSO
spectra with flux scaled to 3060\AA. For the difference spectra
(brighter phase minus fainter phase), the geometric mean composite
difference spectrum is almost the same as the geometric mean
composite spectrum. However, with the arithmetic mean, they are
different, with the slope of the arithmetic mean composite
difference spectrum of $-1.75\pm 0.04$ and the slope of arithmetic
composite spectrum of $-1.42\pm 0.03$. The slope of -1.75 for the
difference spectrum is larger than -2.0 by Wilhite et al. (2005).

In Figure 8, the blue curve is the arithmetic mean composite
spectrum from the SDSS spectra scaled to a value of 1 at 3060\AA.
The ratio of the arithmetic mean composite difference spectrum to
the arithmetic mean composite spectrum is shown as a black curve.
Larger values of this ratio indicate a more variable portion of the
spectrum. We can find smaller values in emission lines of
\civ,\mgii,\hb, etc on the ratio spectrum, showing that the emission
lines are considerably less variable than the continuum. It is
consistent with the intrinsic Baldwin effect for emission lines
(Kinney et al. 90). For rest wavelengths greater than 2500 \AA, the
ratio is flat (near 1), showing that the composite difference
spectrum has the same continuum slope as the composite spectrum.
However, the values blueward of 2500 \AA\ in the ratio spectrum
become larger than 1. Vanden Berk et al. (2004) showed that the
variability dependence appears to flatten around 3000 \AA\ (see
their Figure 13), which is consistent with the flatness in this
dataset around 3000 \AA.

%Since the ratio spectrum is flat excluding the region blueward of
%2500\AA, the steep slope of composite difference spectrum is due to
%the higher value blueward of 2500\AA, which also shows a difference
%in the composite spectra constructed from arithmetic and geometric
%means. The variability for the region blueward of 2500 \AA\ is
%different to that for the region redward of 2500 \AA.

The ratio spectrum from 2500 \AA\ to 3600 \AA\ is flat (near 1),
showing that the composite difference spectra have the same
continuum slope as the composite spectrum. Blueward of 2500 \AA, the
composite difference spectrum (brighter phase minus fainter phase)
has more variance than the average QSO spectrum. Redward of 3600
\AA, the variance becomes slightly greater than 1 and increases
slowly toward longer wavelengths.

Since the ratio spectrum is flat excluding the region blueward of
2500 \AA, the steep slope of the composite difference spectrum is
due to the higher value blueward of 2500 \AA, which also shows a
difference in the composite spectra constructed from both the
arithmetic and geometric means. The variability for the region
blueward of 2500 \AA\ is different from that for the region between
2500 \AA and 3600 \AA, and different from the region redward of 3600
\AA.

Our interpretation of this result is that the lack of variability
from 2500 \AA\ to 3600 \AA\ arises because this spectral region is
dominated by Balmer continuum and Fe II emission.  As pointed out
above, the arithmetic mean is sensitive to spectral features.  The
Fe II region is composed of the many members of the stronger UV
multiplets, and they may completely dominate the continuum emission
(e.g., Wampler 1986).  Since the emission lines in general are shown
to vary less than the continuum in the composite difference
spectrum, this region crowded with line emission and Balmer
continuum (the 3000 \AA\ bump) is likely to have a similarly lower
variability.

The variability for the region blueward of 2500 \AA\ and redward of
3600 \AA\ is due to the accretion disk. During the brighter phase,
the accretion disk becomes hotter and its emission peak would move
to shorter wavelengths (big blue bump), which would lead to larger
variance in the blue spectrum. For the region redward of 3600 \AA\,
the more slowly varying outer regions of the disk would dilute the
variability from the increased flux from the hotter regions,  To the
extent that the increase in variability to the red is significant,
the density of slowly varying emission lines (particularly the
higher Balmer series and optical UV multiplets) is decreasing in
that direction.

The more variance blueward of 2500 \AA\ does not mean that the
spectrum becomes bluer when the spectrum becomes brighter. Due to
accretion power  energy mechanics, it is believed that the
UV-optical continuum can be described by a power-law function.
However, for the different spectrum, it is not the case (see cyan
curves in Figure 2). Sometimes, respect to the ratio spectrum, the
slope would be oppositive if we use a power-law function to fit the
different spectrum.

In Figure 8, we also give the ratio spectra for variable RL QSOs
(red curve), and variable RQ QSOs (green curve). Blueward of 2500
\AA, they are almost the same. And we notice that number of RQ QSOs
are not as many as RL QSOs in our variable sample. There is a small
bump redward of 4000 \AA\ for variable RL QSOs, however, that is not
the case for variable RQ QSOs.

There is no obvious difference in the distribution of the slope
variations for subsamples of RQ and RL QSOs (Figure 4). The color
change with increases in brightness is not different between RL and
RQ QSOs. It implies that the presence of a radio jet does not affect
the slope variability on 10-year timescales. The variability must be
from disk instabilities or some other variation,such as reprocessing
of X-rays and reflection of optical light by the dust (Breedt et al.
2010).

\section{Conclusions}
For the sample of FIRST bright QSOs by White et al. (2000), we
assemble their spectra from SDSS DR7 to investigate the spectral
variability between the spectra from White et al. (2000) and from
SDSS over a long time lag (up to 10 years). There are 312 radio loud
QSOs and 232 radio quiet QSOs in this sample, up to z$\sim$3.5. The
main results are summarized as follows: (1) With two-epoch QSO
variation, it is found that about half of the QSOs appear redder
during their brighter phases, not only for variable RQ QSOs, but
also for variable RL QSOs. (2) We did not find strong correlations
between the slope variation and the continuum flux ratio. (3) The
composite bright-to-faint ratio spectrum is flat for subsamples of
RQ and RL QSOs. There is no obvious difference in slope variations
for subsamples of RQ and RL QSOs. The color change with increases in
brightness is not different between RL and RQ QSOs. It implies that
the presence of a radio jet does not affect the slope variability on
10-year timescales. (4) The arithmetic composite difference spectrum
(bright phase minus faint phase) for our variable QSOs is steep at
blueward of $\sim$ 2500\AA, implying QSO has more variability in the
blue spectrum. The variability for the region blueward of 2500 \AA\
is different to that for the region redward of 2500 \AA.

\section{ACKNOWLEDGMENTS}
We thank an anonymous referee for suggestions that led to
improvements in this paper. This work has been supported by the
National Science Foundations of China (No. 10873010; 11173016;
11233003).

\bibliography{}

\end{document}